\documentclass[twocolumn,showpacs,aps,prb]{revtex4}

\usepackage{graphicx}
\usepackage{amsfonts, amsmath, amstext, amssymb, amsfonts, amsxtra}
\usepackage{braket}
\usepackage{bbm}
\usepackage{bbold}
\usepackage[protrusion=true,expansion=true]{microtype}
\usepackage{wasysym}
\usepackage{cleveref}
\usepackage{color}
\usepackage{ulem}

\newcommand{\im}{{i}}         
\newcommand{\e}{{e}}          
\newcommand{\id}{\mathbb{1}}  

\begin{document}

	\title
	{
		Quantum jumps on Anderson attractors
	}
	
	\author
	{
		I.I.~Yusipov$^{1,4}$, T.V.~Laptyeva$^2$, and M.V.~Ivanchenko$^{3,4}$
	}
	
	\affiliation
	{
		$^{1}$Institute of Supercomputing Technologies, Lobachevsky University, Gagarina Av.\ 23, Nizhny Novgorod, 603950, Russia \\
		$^{2}$Department of Control Theory and Systems Dynamics, Lobachevsky University, Gagarina Av.\ 23, Nizhny Novgorod, 603950, Russia\\
		$^{3}$Department of Applied Mathematics, Lobachevsky State University of Nizhny Novgorod, Gagarina Av.\ 23, Nizhny Novgorod, 603950, Russia
		$^{4}$ Center for Theoretical Physics of Complex Systems, Institute for Basic Science, Daejeon 305-732, Korea
	}


	\begin{abstract}
		In a closed single-particle quantum system, spatial disorder induces Anderson localization of eigenstates and halts wave propagation. The phenomenon is vulnerable to interaction with environment and decoherence, that is believed to restore normal diffusion. We demonstrate that for a class of experimentally feasible non-Hermitian dissipators, which admit signatures of localization in asymptotic states, quantum particle opts between diffusive and ballistic regimes, depending on the phase parameter of dissipators, with sticking about localization centers. In diffusive regime, statistics of quantum jumps is non-Poissonian and has a power-law interval, a footprint of intermittent locking in Anderson modes. Ballistic propagation reflects dispersion of an ordered lattice and introduces a new timescale for jumps with non-monotonous probability distribution. Hermitian dephasing dissipation makes localization features vanish, and Poissonian jump statistics along with normal diffusion are recovered.

	\end{abstract}

	\maketitle
	
{\it Introduction. --} Anderson localization was introduced for a closed disordered quantum system \cite{Anderson1958}, and most of the theoretical studies \cite{Kramer1993,Evers2008,fifty}, as well as experimental observations \cite{Billy2008,Roati2008,Kondov2011,Jen2012} remained in this realm. Although already Anderson pointed it out that a contact to some thermal reservoir ``will actually control the transport processes'' \cite{Anderson1958}, the interest to the issue has been quite limited: Intuition suggests, that decoherence caused by interaction with the environment \cite{book} will undermine destructive interference mechanism of localization. Indeed, early studies   confirmed that dephasing due to dissipation or measurement destroys Anderson localization \cite{Flores1999,Gurvitz2000,Schlagheck2012} (or dynamical localization\cite{Dittrich1990}) and gives way to diffusion; even local measurement proved sufficient for complete delocalization\cite{Gurvitz2000}. 

Recent results, however, elucidate a much richer physics than expected. First, it was demonstrated that even when the asymptotic state is a trivial uniform distribution, the relaxation process manifests heterogeneous dynamics and signatures of metastability \cite{les0}. Second, it was shown that a one-dimensional quantum system with a  Hamiltonian exhibiting  Anderson localization can be driven into a steady state, an ``Anderson attractor'', which retains localization properties \cite{Yusipov2017}. Such an asymptotic state can be engineered with a set of local dissipative operators \cite{DiehlZoller2008, KrausZoller, wolf2009,marcos2012}, the corresponding mechanism is based on the robust spatial phase-structure of Anderson modes \cite{Vershinina2017}.

	 \begin{figure*} [t!]
	 	
	  		{\includegraphics[width=0.95\columnwidth]{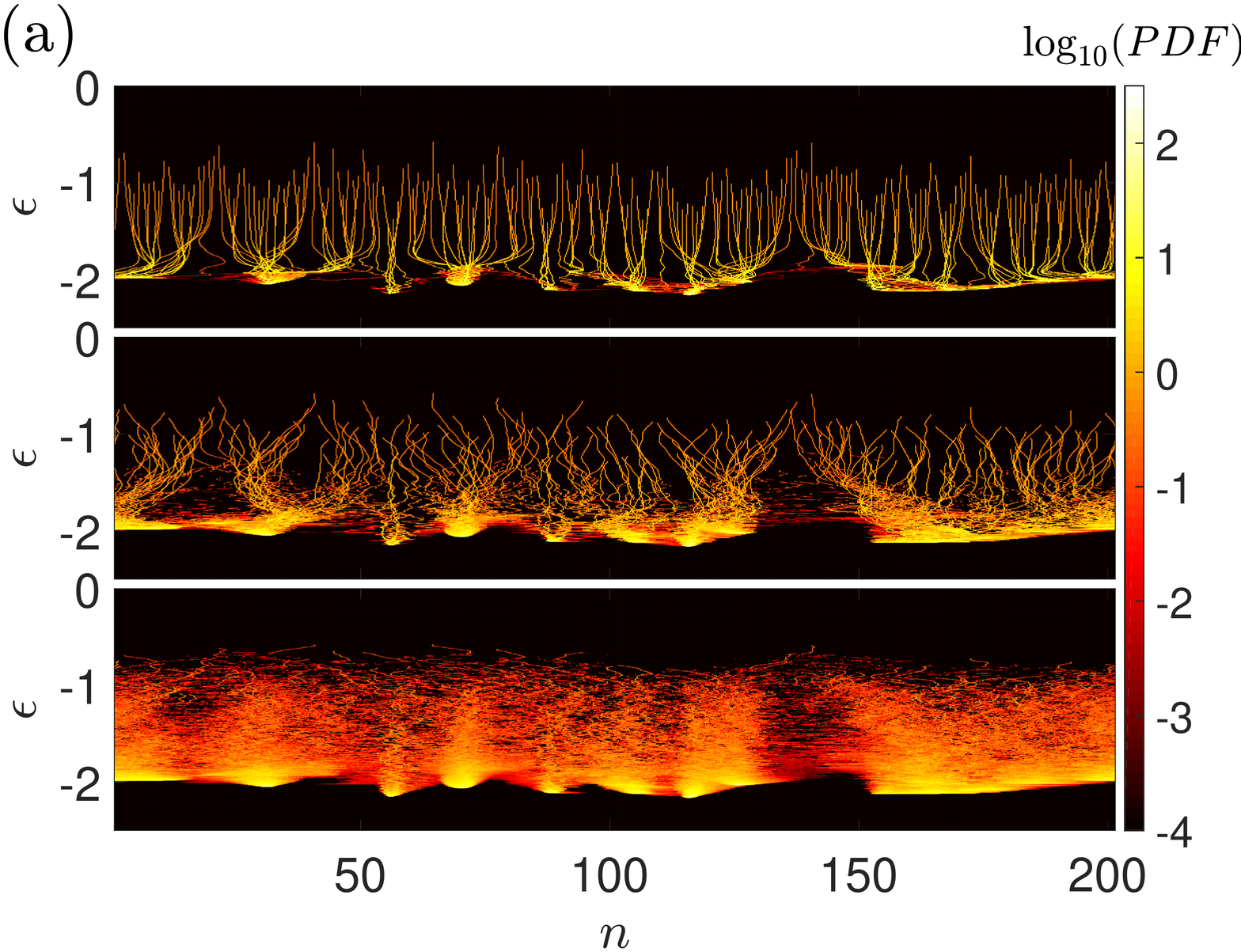}}  
	 	\hfill
	 	 {\includegraphics[width=0.95\columnwidth]{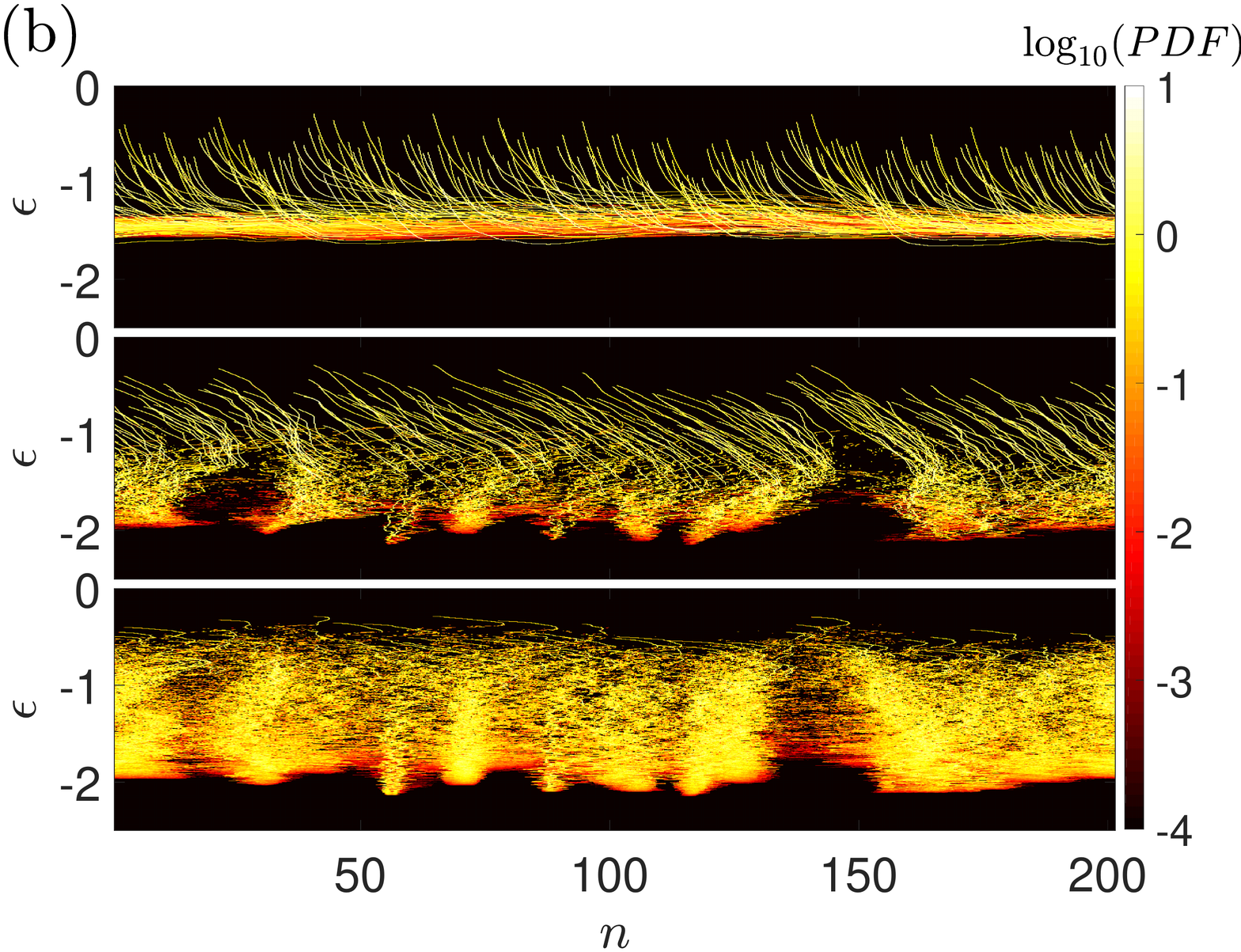}} \\
	 	{\includegraphics[width=0.95\columnwidth]{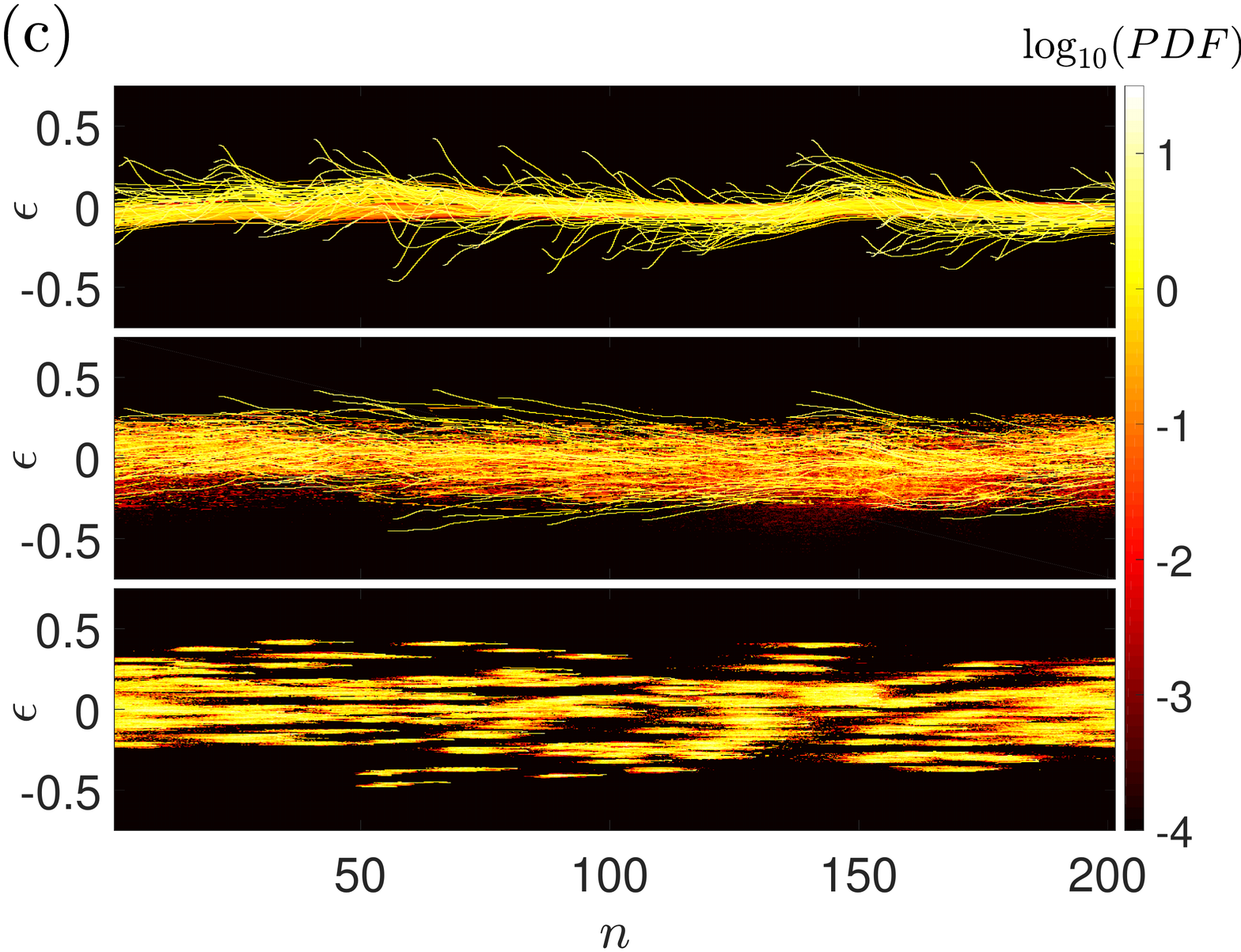}}  
	 	\hfill
	 	{\includegraphics[width=0.95\columnwidth]{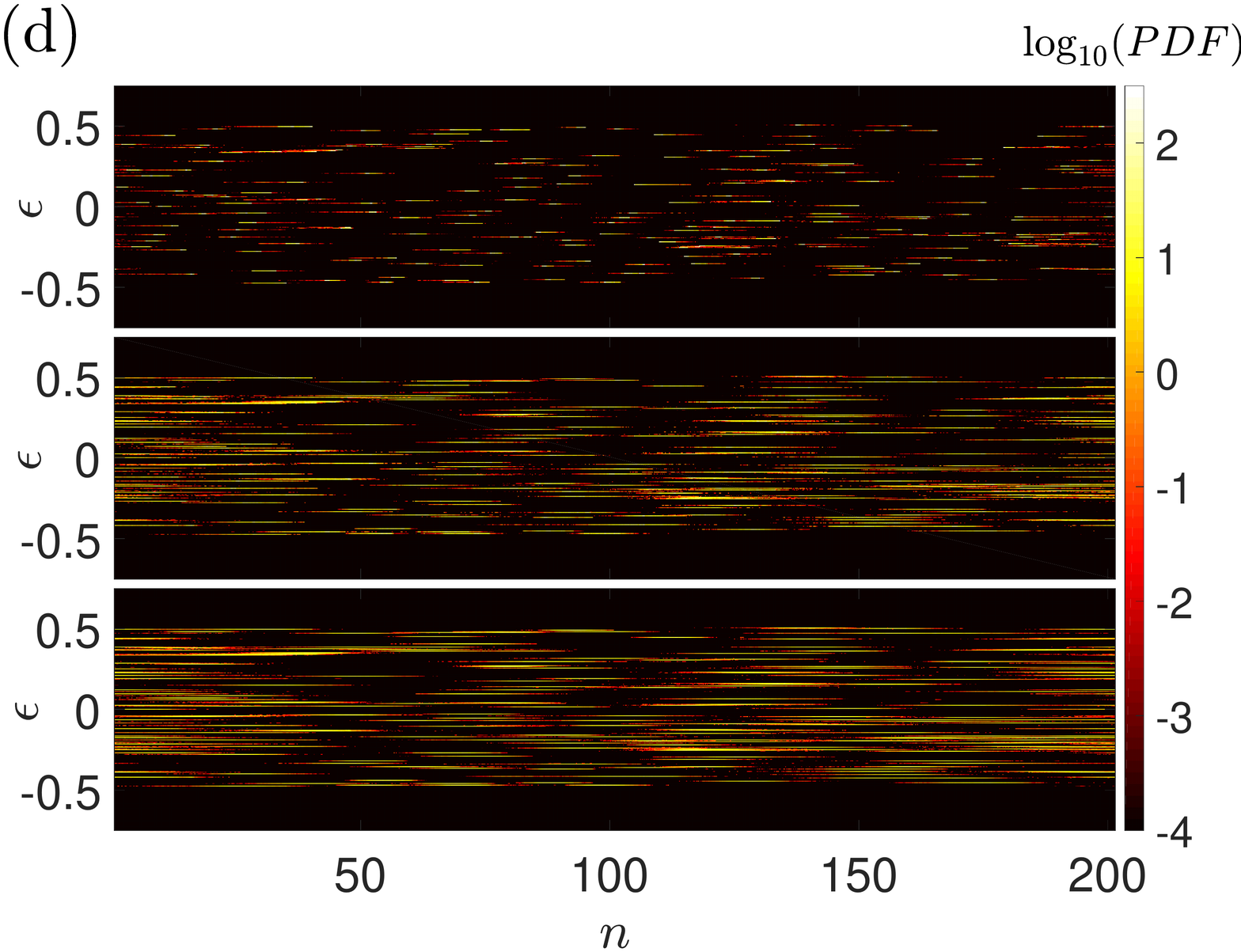}}  
	 	\caption
	 	{
	 		(Color online) Anderson attractors: probability density function (PDF) for quantum trajectories on the mass center -- energy plain in the asymptotic regime for $\gamma=0.1$ (top), $\gamma=0.01$ (middle), $\gamma=0.001$ (bottom). Non-Hermitian dissipator, Eq.(\ref{eq:4}), with phase parameters (a) $\alpha=0$; (b) $\alpha=\pi/4$; (c) $\alpha=\pi/2$; and (d) dephasing dissipator, Eq.(\ref{diss_poletti}). Ensemble averaging is taken over $M_{\mathrm{r}}=10^3$ trajectories, which were propagated up to $T = 10^7$ after relaxation time $t_0=10^3\gamma^{-1}$. Here $W=1, N=200$.
	 	}
	 	
	 	\label{Fig:1}
	 	
	 \end{figure*}

In this Rapid Communication we revisit the Anderson's proposition and investigate the dynamics of a quantum particle on an {\it open} disordered lattice in the asymptotic regime with footprints of localization. Single trajectories are resolved with the quantum Monte-Carlo wave function (quantum jump) method \cite{dali,zoller1992,Plenio1998}. We demonstrate that they are shaped by the competition of (i) diffusion, built of sticking and intermittent jumps between localization centers, and (ii) ballistic propagation inherited from the dark states of the disorder-free system. Controlling the phase properties of local dissipators allows for switching between diffusive and ballistic regimes, and varying the direction and speed of the latter. Statistics of quantum jumps is non-Poissonian, reflecting an interplay between disorder and dissipation. In case of dephasing dissipation, localization features vanish, and Poissonian jump statistics along with normal diffusion are restored.    
	
	{\it Model. --} The open Anderson system is described by the Lindblad master equation \cite{book, Alicki1987},
	\begin{align}
		\dot{\varrho} = \mathcal{L}(\varrho) = -\im [H,\varrho] + \mathcal{D}(\varrho).
		\label{lindblad}
	\end{align}
	The first term on the r.h.s.\ captures the unitary evolution of the system
	governed by a single-particle Anderson Hamiltonian $H$:
	\begin{align}
		H = \sum_j \epsilon_j b_j^{\dagger}b_j -(b_{j}^{\dagger} b_{j+1} + b_{j+1}^{\dagger} b_{j}), 
		\label{hamiltonian}
	\end{align}
	where $\epsilon_j\in\left[-W/2, W/2\right]$ are random uncorrelated on-site energies, $W$ is the 
	disorder strength, $b_j$ and $b_j^{\dagger}$ are the annihilation and creation operators of a boson on the $j$-th site. The eigenvalues of the Hamiltonian are $E_q \in \left[-2-W/2, 2+W/2 \right]$, while the respective eigenstates, $A_j^{(q)}$,  are exponentially localized with length\cite{Thouless1979} $\xi_E \approx 24(4-E^2)/W^2$  . Periodic boundary conditions are assumed, $\varrho_{0} = \varrho_{N+1}$.
	
	The term in the Lindblad equation that describes dissipation,
	\begin{align}
		\mathcal{D}(\varrho) = \sum_{j=1}^{S} \gamma_{j}(t) \left[V_j\varrho V^\dagger_j - \frac{1}{2}\{V^\dagger_jV_j,\varrho\}\right],
		\label{dissipator}
	\end{align}
	involves the set of $S$ operators, $\{V_j\}_{1,...,S}$, which capture action of the environment on the system.

	\begin{figure*} [t!]
				{\includegraphics[width=0.95\columnwidth]{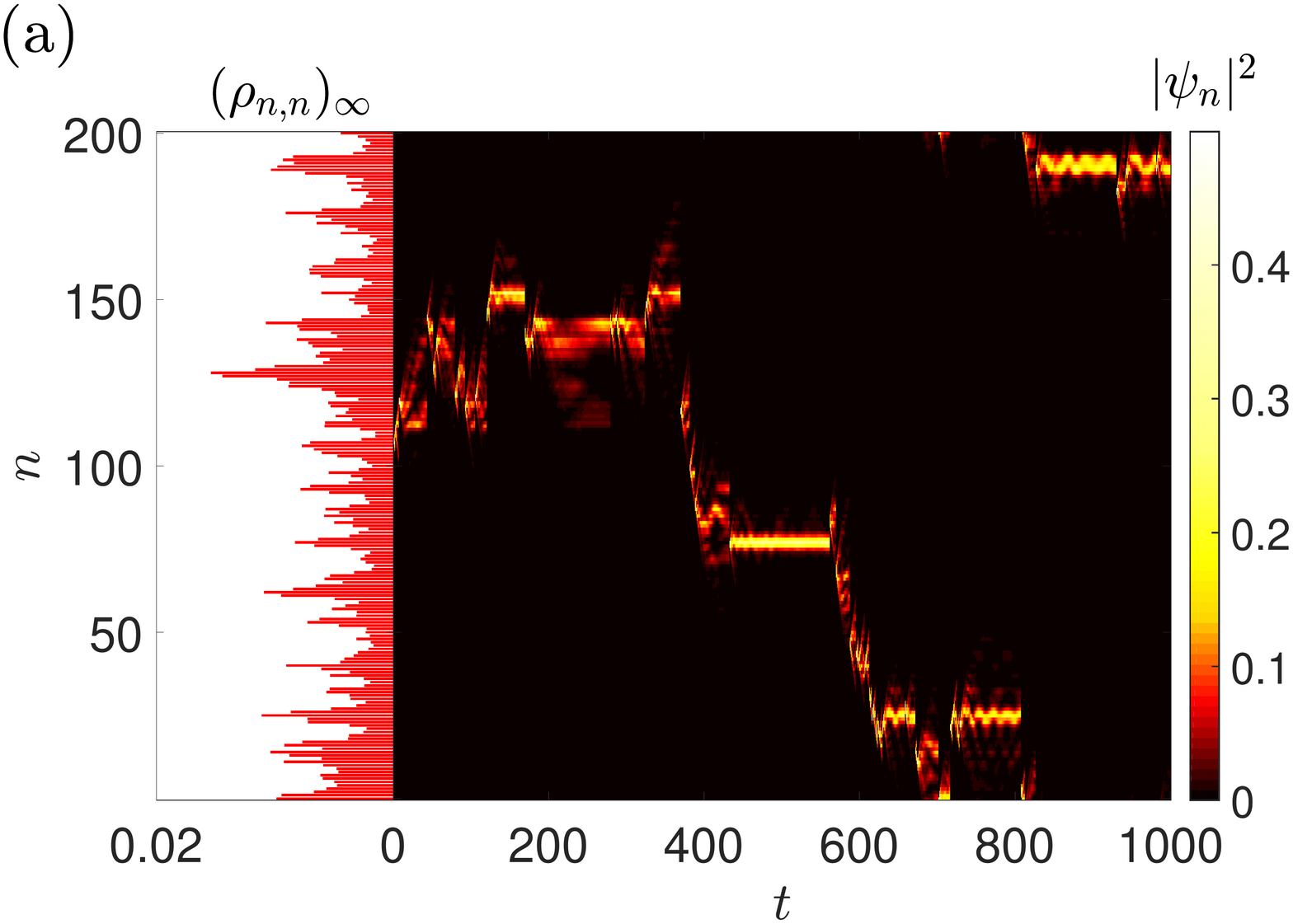}} 
		\hfill
			 {\includegraphics[width=0.95\columnwidth]{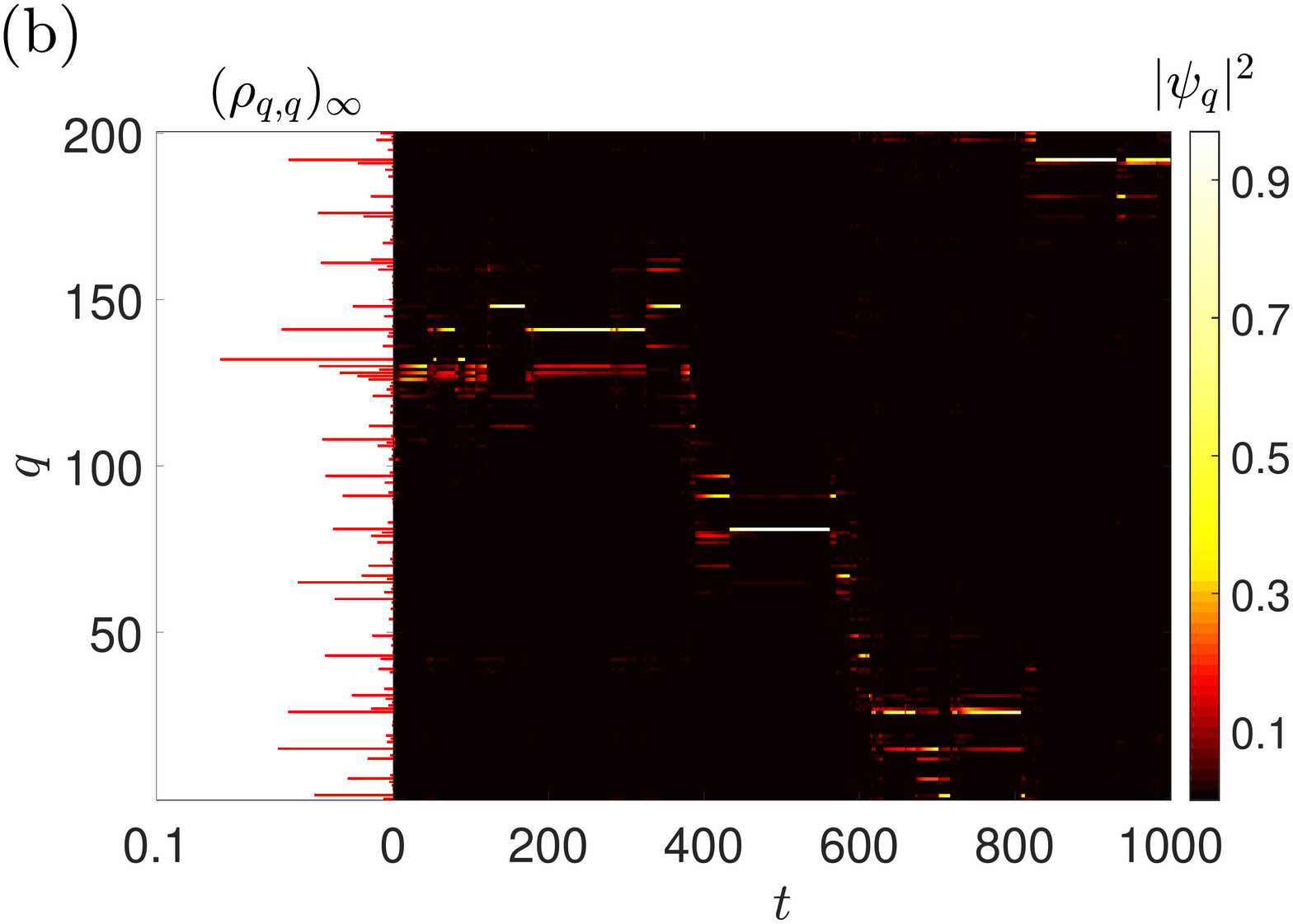}} \\
		 	{\includegraphics[width=0.95\columnwidth]{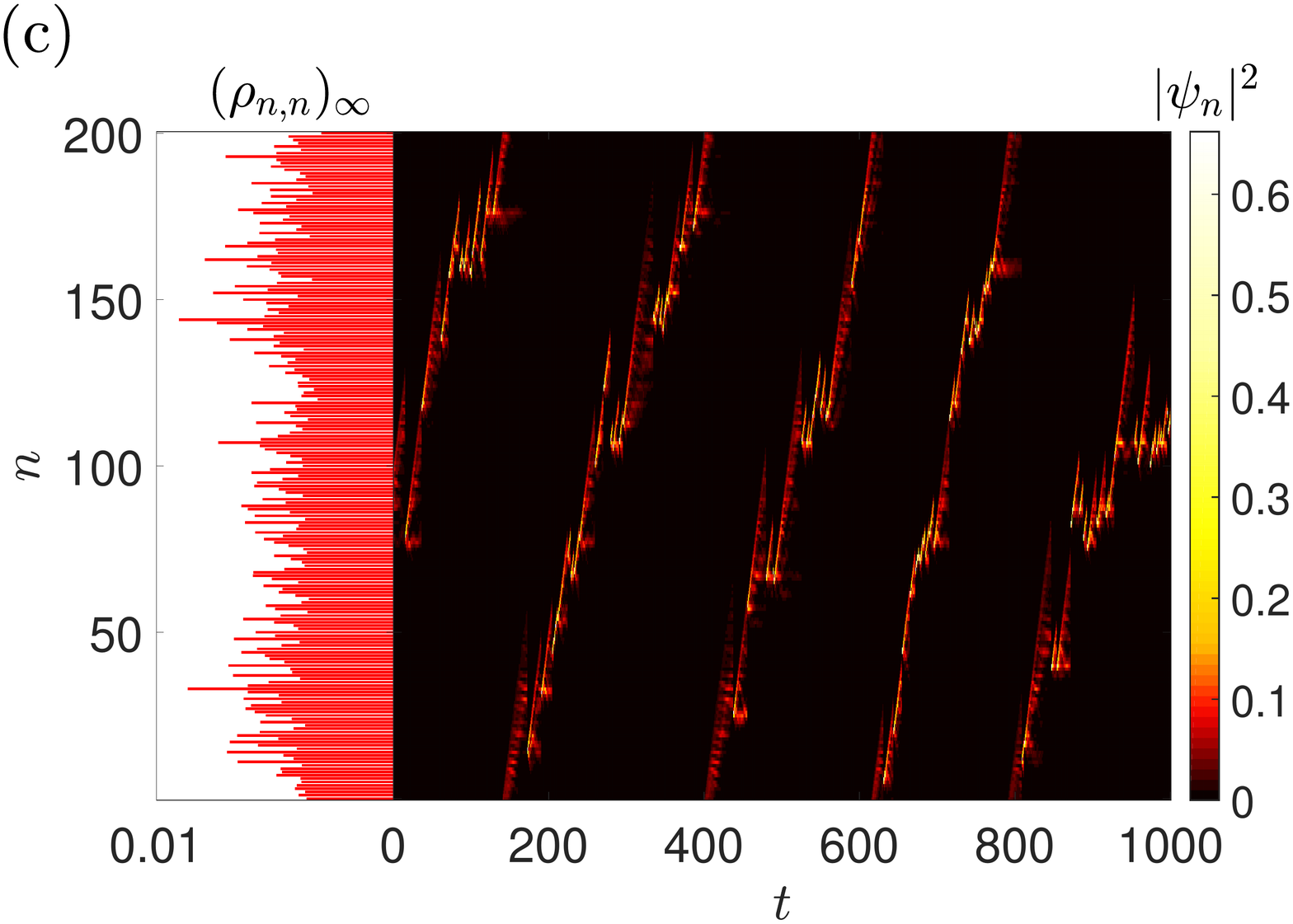}}
		\hfill
			{\includegraphics[width=0.95\columnwidth]{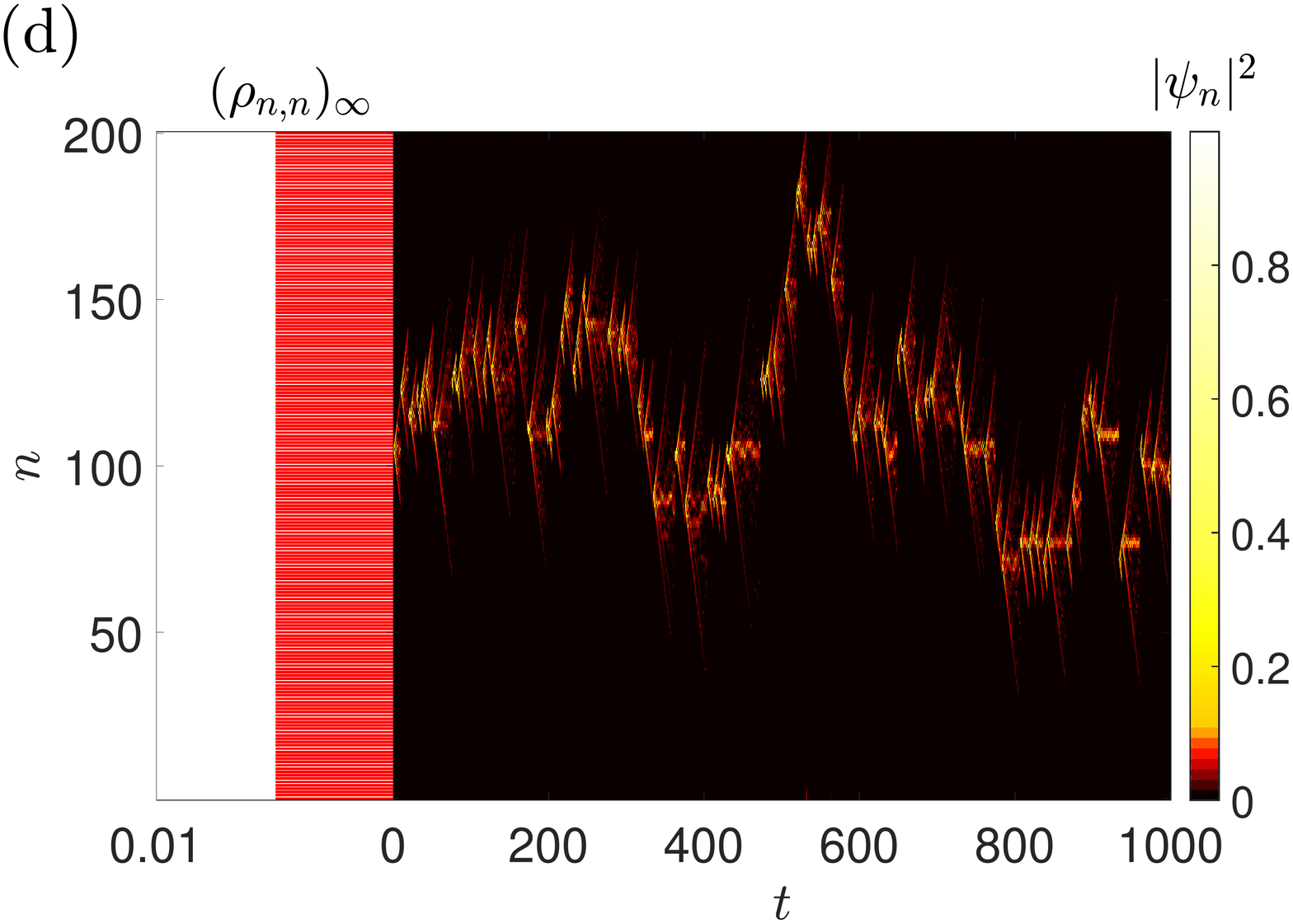}}  
		\caption
		{
			(Color online) Asymptotic density matrix diagonal elements (left-side panels) and single quantum trajectories on Anderson attractors (main panels): non-Hermitian dissipators Eq.(\ref{eq:4}) with (a) $\alpha = 0$, direct space; (b) $\alpha = 0$, Anderson basis, the modes are ordered by the center of mass coordinate; (c) $\alpha = \pi/4$, direct space; and (d) dephasing dissipators Eq.(\ref{diss_poletti}), direct space. Here $W=1$, $\gamma=0.1$, and $N=200$.
		}
		
		\label{Fig:2}
		
	\end{figure*}

	The structure of asymptotic density matrix, which can be defined under some conditions \cite{Alicki1987} as $\varrho_{\infty} = \lim_{t \rightarrow \infty} \e^{\mathcal{L}t} \varrho_0$ (for all $\varrho_0$) highly depends on the Hamiltonian and form of dissipative operators. This asymptotic matrix is found as a kernel of the  Lindblad generator
	in Eq.~(\ref{lindblad}), $\mathcal{L}(\varrho_{\infty}) = 0$.

We consider local non-Hermitian dissipators \cite{DiehlZoller2008, KrausZoller, marcos2012, Bardyn2013, Barreiro2010,Kienzler2015, Vorberg2013}
	 \begin{align}
	 	V_j=(b_{j}^{\dagger} + e^{i \alpha}b_{j+1}^{\dagger})(b_{j} - e^{-i\alpha}b_{j+1}),
	 	\label{eq:4}
	 \end{align}
which produce non-trivial asymptotic states featuring localization, coined ``Anderson attractors'' \cite{Yusipov2017,Vershinina2017}.  The dissipators are parametrized by\cite{implementation} $\alpha$, making them phase-selective. For example, when $\alpha=0$, the operator tries  to synchronize the dynamics on the $j$ and $j+1$ sites, by constantly recycling anti-symmetric out-of-phase mode into the symmetric in-phase one; the effect of $\alpha=\pi$ is the opposite. More generally, a zero-disorder eigenstate  $\psi_{j}=e^{i k j}/\sqrt{N}, \ k=2\pi q/N, q=-N/2\ldots N/2$ is a dark state of the dissipators for $\alpha=k$. As Anderson modes inherit spatial phase properties from the seeding plain waves \cite{ishii}, asymptotic states of the open disordered lattice are dominated by a respective part of the Anderson spectrum, controlled by\cite{Yusipov2017,Vershinina2017} $\alpha$.      	
	
To provide with a reference case, we also consider dephasing dissipators \cite{les0,Fischer2016,Levi2016,Everest2017}
	\begin{align}
		V_j =b_{j}^{\dagger} b_{j},
		\label{diss_poletti}
	\end{align}
which universally produce a trivial asymptotic density matrix $\varrho_{\infty} = \id/N$. We also assume time-independent and identical coupling to dissipation channels, $\gamma_j(t)=\gamma$. 
	
Although the asymptotic density matrix, $\varrho_{\infty}$, describes a statistical distribution of single quantum trajectories, it lacks information on their {\it microscopic dynamics} in the asymptotic regime, and, therefore, is not sufficient for our purpose. We employ quantum Monte-Carlo wave function (quantum jump) method to unravel  the deterministic  equation (\ref{lindblad}) into an ensemble of quantum trajectories \cite{dali,zoller1992,Plenio1998}. It recasts the evolution of the model system in terms of pure states and wave function, $\psi(t)$, governed by an effective non-Hermitian Hamiltonian,
	 \begin{align}
	 	\tilde{H} = H -\frac{i}{2}  \sum_j  V^\dagger_j V_j,
	 	\label{jumps_hamiltonian}
	 \end{align} 
	 and random jumps induced by dissipators $V_j$. 
		In all experiments we generate up to $M_{\mathrm{r}}=10^3$ different trajectories, leave $t_0 = 10^3\gamma^{-1}$ time for relaxation towards an asymptotic state, and follow the dynamics for up to $T = 10^7$. 

	\begin{figure} [t!]
	{\includegraphics[width=0.95\columnwidth]{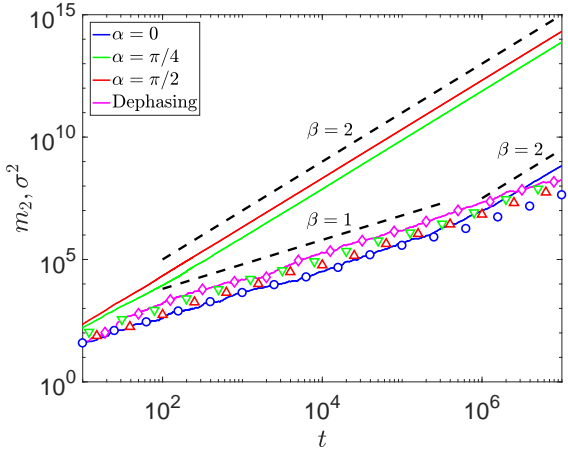}}
	
		\caption
		{
			(Color online) Evolution of the second moment $m_2(t)$ (solid lines) and mean square displacement $\sigma^2(t)$ (symbols) for the non-Hermitian dissipators with $\alpha=0$ (blue), $\alpha=\pi/4$ (green); $\alpha=\pi/2$ (red); and dephasing dissipation (magenta). Black dash-dotted lines indicate power laws $\propto t^\beta$. Here $W=1$, $\gamma=0.1$, $N=200$. 
		}
		
		\label{Fig:3}
		
	\end{figure}

	 We start with investigating the fine structure of asymptotic states in dependence on the dissipation rate, $\gamma$. An ensemble of quantum trajectories yields the probability density function (PDF) on the mass center and energy expectations plain $\{n(t), \epsilon(t)\}$, 
	 \begin{align}
	 	n(t)&=\sum_k\langle\psi(t)|b^\dagger_jb_j|\psi(t)\rangle, \\
	 	\epsilon(t)&=\langle\psi(t)|H|\psi(t)\rangle.
	 	\label{mc_energy}
	 \end{align}
	 where the mass center is calculated with regard to periodic boundaries. (Instructively, distributions of asymptotic diagonal elements, $(\varrho_{n,n})_\infty$, proved to be weakly dependent on $\gamma$.) Fig. \ref{Fig:1} presents a typical picture for a fixed realization of disorder. For $\alpha=0$, the trajectories bundle up about localization centers, connected by a web of transitions; convergence to bundles and their compactness weaken with dissipation rate (Fig. \ref{Fig:1}(a), from top to bottom). Non-zero $\alpha=\pi/4$ introduces a pronounced skew in trajectories (Fig. \ref{Fig:1}(b)); ultimately, localization centers get invisible for $\alpha=\pi/2$, (Fig. \ref{Fig:1}(c)). Note that localization in the energy persists, varying from the lower edge of the Anderson spectrum ($\alpha=0$) to its middle ($\alpha=\pi/2$). In contract, dephasing dissipation leads to a random structure, spanned over the whole range of disorder energies (Fig.\ref{Fig:1}(d)). 

Next, we follow single quantum trajectories, $\psi(t)$, that evolve under Eq.(\ref{jumps_hamiltonian}), and compare them against the profile of the asymptotic state, $(\rho_{n,n})_\infty$ (Fig.\ref{Fig:2}). In case of $\alpha=0$, we observe an intermittent dynamics of long sticking about localization centers and rapid transitions between them (Fig.\ref{Fig:2}(a)). Recasting the picture in the Anderson basis reveals that sticking occurs at the Anderson modes, which dominate the asymptotic state (Fig.\ref{Fig:2}(b)). Non-zero phase parameter of dissipator dramatically changes the dynamics: although sticking about localization centers is traceable, it now becomes overlaid with ballistic propagation, see Fig.\ref{Fig:2}(c) for $\alpha=\pi/4$. Lastly, the dephasing dissipation leads to random jumps that lack any spatio-temporal structure (Fig.\ref{Fig:2}(d)).

To quantify the quantum particle propagation we follow its center of mass, $n(t)$, calculating the ensemble averaged second moment of displacement, $m_2(t)=\left\langle [n(t)-n(t_0)]^2\right\rangle$, average velocity, $v~=~\left\langle[n(T)-n(t_{0})]/(T-t_{0})\right\rangle$, and mean square displacement from an average ballistic trajectory, $\sigma^2(t)~=~\left\langle[n(t)-n(t_{0})-v\cdot(t-t_0)]^2\right\rangle$.

	\begin{figure} [t!!!!]
		{\includegraphics[width=0.95\columnwidth]{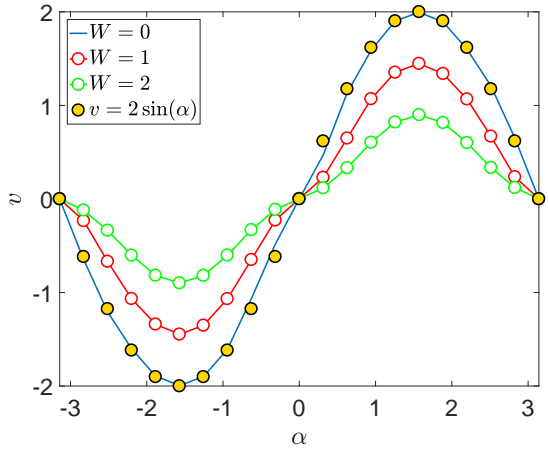}}
		\caption
		{(Color online) Velocity of the wave packet propagation in dependence on the phase of dissipator, $v(\alpha)$ . Averaging is taken over $M_{\mathrm{r}}=10^3$ different trajectories, propagated to $T=10^7$. The other parameters are $\gamma=0.1$, $N=200$.
		}
		\label{Fig:4}
	\end{figure}

	\begin{figure} [t!]
	{\includegraphics[width=0.95\columnwidth]{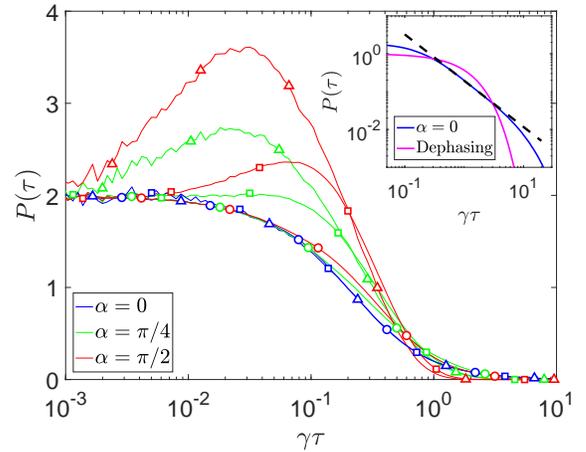}} 
		\caption
		{(Color online) Probability distribution of times between quantum jumps. Non-Hermitian dissipator, Eq.(\ref{eq:4}), with $\alpha = 0$ (blue), $\alpha = \pi/4$ (green), and $\alpha = \pi/2$ (red). Markers correspond to dissipation rate: $\gamma=0.1$ (circles), $\gamma=0.01$ (squares), $\gamma=0.001$ (triangles). Inset: $\alpha=0$ (blue) and dephasing dissipator (magenta), black dashed line indicates a power-law $P(\tau)\propto \tau^{-1}$, here $\gamma=0.1$. The other parameters are $W=1$, $N=200$. 
		}
		\label{Fig:5}
	\end{figure}

Remarkably, evolution of the second moment manifests different power laws, $m_2(t)\propto t^\beta$, with normal diffusion, $\beta\approx1$, for $\alpha=0$ and dephasing dissipation, and ballistic spreading, $\beta\approx2$, for $\alpha=\pi/4$ and $\alpha=\pi/2$ (Fig.\ref{Fig:3}, solid lines). At the same time, the squared standard deviation from an average ballistic trajectory demonstrates an accompanying diffusion, $\sigma^2(t)\propto t^\beta$, $\beta\approx1$ (Fig.\ref{Fig:3}, symbols). Noteworthy, the normal diffusion for $\alpha=0$ is taken over by ballistic propagation at asymptotically large times, $t\sim10^6\ldots10^7$, which appears to be a finite size effect. 

The switch from diffusive to ballistic propagation for non-zero $\alpha$ can be understood as an interplay between disorder and dissipation. As we already pointed it out, dissipation selects Anderson modes from a particular part of the spectrum, and they borrow spatial phase properties of the zero-disorder plain wave  eigenstates, with wave numbers\cite{ishii,Vershinina2017} $k\approx\alpha$. Overlapping in space  (Figs.\ref{Fig:1} and \ref{Fig:2}), the exponentially localized modes interact due to dissipative coupling. It enables directed propagation of a quantum wave packet with characteristic velocity, sensitive to a preferred wave number. 

Extensive numerical simulations reveal the dependence of the wave packet velocity on the phase of dissipation, $v(\alpha)$, Fig.\ref{Fig:4}. In the disorder-free array, $W=0$, it is given by the group velocity of plain waves, $v(\alpha)=v_{group}(k)|_{k=\alpha}=2\sin\alpha$, the dark states when $k=\alpha$. Disordered array manifests the functional dependence of the sine shape, the magnitude decreasing for greater disorder (Fig.\ref{Fig:4}).        
	
	To get a deeper insight into statistic of single quantum trajectories, we study probability distributions of time intervals between the jumps, $P(\tau)$. First, we look into the case $\alpha=0$, where localization is most pronounced, and the trajectory displays long-time sticking at the dominant Anderson modes (Fig.\ref{Fig:2}(a,b)). It turned out that such intermittency leaves a footprint on inter-jump time distribution, seen as a power law interval, $P(\tau)\sim\tau^{-1}$, in a drastic difference to the Poisson statistics for dephasing dissipation, $P(\tau)\sim e^{-\tau}$ (Fig.\ref{Fig:5}, inset). 
	
	Additional features arise in dependence on $\alpha$ (Fig.\ref{Fig:5}, main part). For $\alpha=0$, when propagation is diffusive, the distribution scales with the dissipation rate, $\gamma$, such that $P(\gamma\tau)$ remains almost the same. This is quite natural as the only temporal scale is given by $\gamma$-dependent quantum jumps between different Anderson modes. The picture changes for non-zero $\alpha$ with the onset of ballistic spreading. While for moderate dissipation rate, $\gamma=0.1$, the distributions for $\alpha=\pi/4$ and $\alpha=\pi/2$ are not much different from the previous, weak dissipation, $\gamma=0.01, 0.001$, gives a pronounced maximum. Arguably, this is a signature of a new timescale, a characteristic passage time of a wave packet across an Anderson mode with an average propagation speed, as determined by the phase $\alpha$ and disorder strength $W$(Fig.\ref{Fig:4}). It limits sticking time at Anderson modes from above, more substantially for smaller $\gamma$, when the other timescale increases.

{\it Conclusions. --} A quantum particle in an open Anderson system can manifest a complex behavior determined by the interplay between disorder and dissipation. For a class of experimentally feasible non-Hermitian dissipators with an adjustable phase property the asymptotic states -- Anderson attractors -- are built of Anderson modes from a narrow part of spectrum.  Single trajectories, resolved with the quantum Monte-Carlo wave function (quantum jump) method, participate in (i) normal diffusion with sticking and intermittent jumps between localization centers, overlaid with (ii) ballistic propagation, dictated by the dark states of the disorder-free system. Controlling the phase parameter of local dissipators, one obtains diffusive or ballistic propagation, the latter reproducing dispersion of an ordered lattice to some extent. In diffusive regime, statistics of quantum jumps is non-Poissonian and has a power-law interval, a footprint of intermittent locking in Anderson modes. Ballistic propagation introduces a new timescale for jumps and limits sticking times, resulting in the non-monotonous probability distribution of times between jumps.

Our  findings are relevant to a broad range of localizing systems, where non-trivial asymptotic states might be possible for certain classes of dissipation, like quasiperiodic (Aubry-Andre) potentials\cite{Roati2008} and systems with many-body localization \cite{Fischer2016,Levi2016,Everest2017}. Dissipative effects in the presence of interactions that yield subdiffusion \cite{Ivanchenko2014,Frahm2016,Ivanchenko2017} or ballistic spreading \cite{Khomeriki2012,Frahm2015} in a few particle case, is yet another intriguing venue for future investigation. 

{\it Acknowledgments. --} This work was supported by the Russian Science Foundation grant No.\ 15-12-20029. 

\end{document}